\pdfoutput=1

\def\year{2021}\relax
\documentclass[letterpaper]{article} % DO NOT CHANGE THIS
\usepackage{aaai21}  % DO NOT CHANGE THIS
\usepackage{times}  % DO NOT CHANGE THIS
\usepackage{helvet} % DO NOT CHANGE THIS
\usepackage{courier}  % DO NOT CHANGE THIS
\usepackage[hyphens]{url}  % DO NOT CHANGE THIS
\usepackage{graphicx} % DO NOT CHANGE THIS
\usepackage{natbib}  % DO NOT CHANGE THIS AND DO NOT ADD ANY OPTIONS TO IT
\usepackage{caption} % DO NOT CHANGE THIS AND DO NOT ADD ANY OPTIONS TO IT
\usepackage{color}
\usepackage{float}
\usepackage{booktabs}

\title{Comparison and Analysis of Deep Audio Embeddings for \\ Music Emotion Recognition}
\author{

    Eunjeong Koh, Shlomo Dubnov
    \\
}

\affiliations{
    Music Department\\
    University of California, San Diego
    La Jolla, CA 92039\\
    \{eko, sdubnov\}@ucsd.edu

}
\begin{document}

\maketitle

\begin{abstract}
Emotion is a complicated notion present in music that is hard to capture even with fine-tuned feature engineering. In this paper, we investigate the utility of state-of-the-art pre-trained deep audio embedding methods to be used in the Music Emotion Recognition (MER) task. Deep audio embedding methods allow us to efficiently capture the high dimensional features into a compact representation. We implement several multi-class classifiers with deep audio embeddings to predict emotion semantics in music. We investigate the effectiveness of $L^3$-Net and VGGish deep audio embedding methods for music emotion inference over four music datasets. The experiments with several classifiers on the task show that the deep audio embedding solutions can improve the performances of the previous baseline MER models. We conclude that deep audio embeddings represent musical emotion semantics for the MER task without expert human engineering.
\end{abstract}

\section{Introduction}

% \shlomo{please put your text using this format. many thanks!}

% Emotional information in music gives an essential step for music indexing and recommendation task. Previous Music Emotion Recognition (MER) studies explore the sound components that can be used to analyze emotions, such as duration, pitch, velocity, and melodic interval. Those representations are high-level acoustic features based on the domain knowledge \cite{wang2015histogram,madhok2018sentimozart,chen2016scheme,lin2013exploration}.

It is an essential step for music indexing and recommendation tasks to understand emotional information in music. Previous Music Emotion Recognition (MER) studies explore sound components that can be used to analyze emotions such as duration, pitch, velocity, and melodic interval. Those representations are high-level acoustic features based on domain knowledge \cite{wang2015histogram,madhok2018sentimozart,chen2016scheme,lin2013exploration}.

%%%%%%%%%%%
%Learning acoustic features based on domain knowledge needs to be accompanied by human knowledge in defining which features to be extracted.

%%%%%%%%%%%
% However, designing acoustic features for emotion requires significant domain knowledge.
%%%%%%%%%%%

%Those acoustic features are important information for understanding music and emotions.
%However, there is a limitation that human resources cannot be always involved in the pre-processing of large amounts of new data. 
%%%%%%%%%%%
% Human resources cannot be always involved in the pre-processing of large amounts of new data. Furthermore, existing emotion-related features are not generalizable across different datasets because such features are fine-tuned for the target dataset based on the music domain expertise \cite{panda2018novel}.
%%%%%%%%%%%
%In order to solve the difficulty of this problem, we decide to take a closer look at the research that analyzes and extracts the unique features of sound.

%%%%%%%%%%%

Relying on human expertise to design the acoustic features for pre-processing large amounts of new data is not always feasible. Furthermore, existing emotion-related features are often fine-tuned for the target dataset based on music domain expertise and are not generalizable across different datasets \cite{panda2018novel}.

% Past research shows that 
%For computing emotion information without human involvement, previous studies utilize neural networks to extract and analyze the salient semantics of acoustic features.
% On the other hand, deep neural networks allow us to learn useful domain-agnostic representations, or deep embeddings, from raw audio input data. It is reported that deep embeddings frequently outperform hand-crafted feature representations in other signal processing problems such as Sound Event Detection (SED) and video tagging task \cite{DCASE2019Task4}.

Advancement in deep neural networks now allows us to learn useful domain-agnostic representations, known as deep audio embeddings, from raw audio input data with no human intervention. Furthermore, it has been reported that deep audio embeddings frequently outperform hand-crafted feature representations in other signal processing problems such as Sound Event Detection (SED) and video tagging task \cite{wilkinghoff2020open,DCASE2019Task4}.

%For example, some recent works on Sound Event Detection (SED) explore deep representations which are the method to capture information related to novel sound-feature aspects, dependent on sound context and audio features.
%There are multiple examples of deep embedding solutions such as Look, Listen, and Learn network ($L^3$-Net) \cite{cramer2019look} and VGGish \cite{hershey2017cnn,jansen2017large,jansen2018unsupervised} which are audio embeddings to predict video tags from the Youtube-8M dataset \cite{abu2016youtube}.

% The power of deep audio embeddings is to automatically identify predominant aspects in the data at scale. Specifically, a Mel-based Look, Listen and Learn network ($L^3$-Net) embedding method matches state-of-the-art performance on the SED task recently \cite{cramer2019look}. Based on a sufficient amount of training data (around 60M training samples) and training design choices, Cramer et al. reveal the efficacy of $L^3$-Net audio embeddings for detecting novel sound features of each audio clips \cite{cramer2019look}. These findings reflect on using an optimal pre-trained $L^3$-Net model.

The power of deep audio embeddings is to automatically identify predominant aspects in the data at scale. Specifically, the Mel-based Look, Listen, and Learn network ($L^3$-Net) embedding method recently matched state-of-the-art performance on the SED task \cite{cramer2019look}. Using a sufficient amount of training data (around 60M training samples) and carefully designed training choices, Cramer et al. were able to detect novel sound features in each audio clip using the $L^3$-Net audio embeddings \cite{cramer2019look}. Cramer et al. released their optimal pre-trained $L^3$-Net model which can now be extended to new tasks.

% In this paper, we use the deep audio embeddings of $L^3$-Net and VGGish for representing musical emotion semantics. VGGish is also a type of deep audio embedding method based on a VGG-like structure to predict video tags from the Youtube-8M dataset \cite{abu2016youtube,hershey2017cnn,jansen2017large,jansen2018unsupervised}.
%and test the trustworthiness of deep music representations, considering musical emotion semantics. 
% We use those two pre-trained models originally designed for the SED task in order to compute the music emotion information. For evaluating the performance of the embedding methods, we extract deep audio embeddings over four different music emotion datasets, and then we train several classification models with the embeddings for evaluating their efficacy for the MER task.
%show the impact of deep audio embeddings for the MER task with several classifiers.
% The results show that the embedding methods provide an effective knowledge transfer mechanism between SED and MER domains without any additional training samples. More importantly, deep audio embedding methods allow us to identify proper sound features for emotion prediction without difficult feature engineering with domain expertise. Our study helps to transmit the audio-domain knowledge from the SED task into that of the MER task.

In this paper, we compare and analyze the deep audio embeddings, $L^3$-Net and VGGish, for representing musical emotion semantics. VGGish is also a type of deep audio embedding method based on a VGG-like structure trained to predict video tags from the Youtube-8M dataset \cite{abu2016youtube,hershey2017cnn,jansen2017large,jansen2018unsupervised}. We repurpose the two deep audio embeddings, originally designed for the SED task, to the task of MER. In evaluating the performance of the embedding methods over four different music emotion datasets, we use the embeddings in several classification models and evaluate their efficacy for the MER task on each dataset.

Our results show that the embedding methods provide an effective knowledge transfer mechanism between SED and MER domains without any additional training samples. More importantly, the deep audio embedding does not require expert human engineering of the sound features for the emotion prediction task. Our study reveals that audio-domain knowledge from the SED task can be extended to the MER task.

\section{Related Work}

% \subsection{Transfer Learning for Music Emotion Recognition}

% Following successes in the fields of Computer Vision and Natural Language Processing, 
% Deep learning approaches have gained increasing interest in the MIR field. 
% \jason{Explain MER first. And then audio embedding as a candidate solution.}

\begin{figure}[htbp]
  \centering
  \includegraphics[width=\columnwidth]{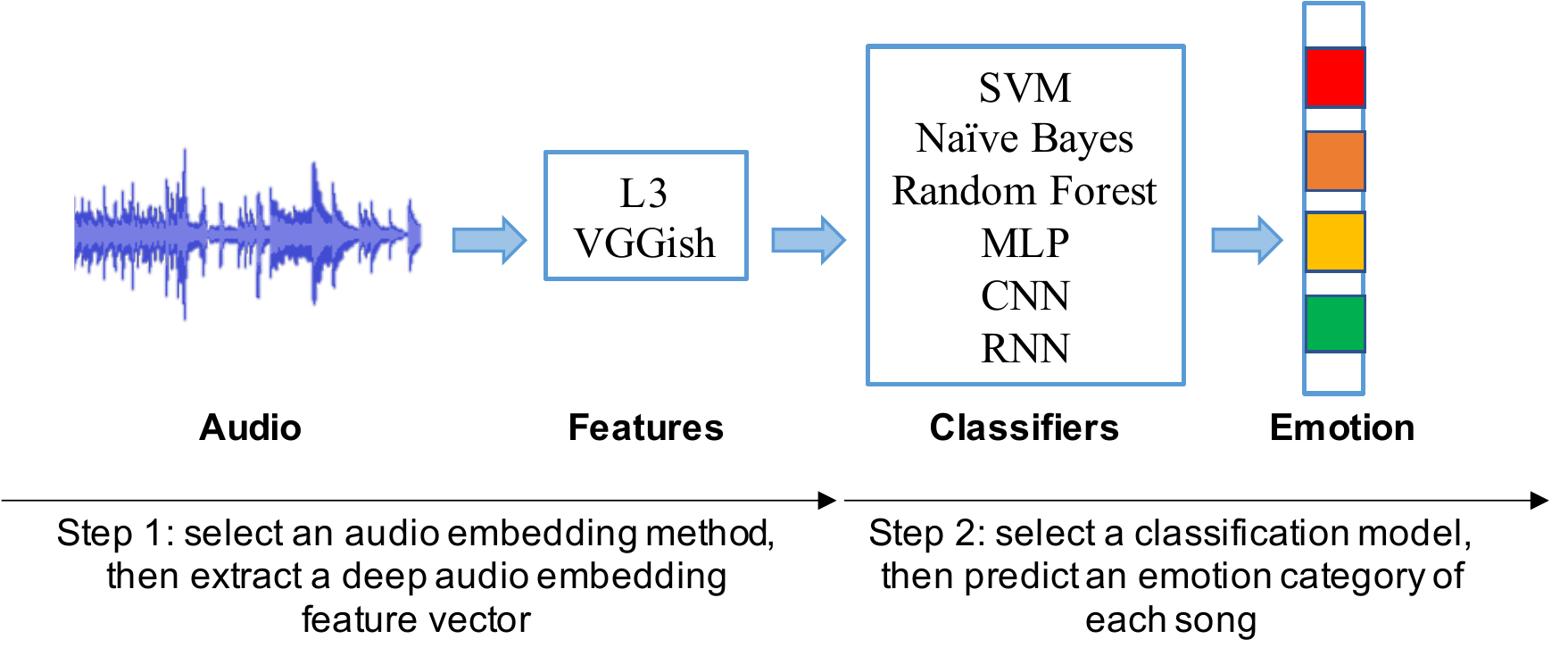}
\caption{\textbf{The Proposed Workflow.} The figure shows the proposed approach using deep audio embeddings for the MER task.}
\label{fig:tl_workflow}
\end{figure}

% The goal of the MER task is to automatically recognize the emotional information conveyed in music \cite{kim2010music}. Although there are many studies in the MER field \cite{soleymani20131000,yang2012machine,yang2018review}, it is a complex process to compare features and performances of the studies because of the technical differences in data representation, emotion labeling, and feature selection algorithm. In addition to that, the proposed methods in different studies are difficult to reproduce since most of them use different public datasets or private datasets with small amounts of music clips and different level of features.

% \shlomo{

One of the goals of the MER task is to automatically recognize the emotional information conveyed in music \cite{kim2010music}. Although there are many studies in the MER field \cite{soleymani20131000,yang2012machine,yang2018review}, it is a complex process to compare features and performances of the studies because of the technical differences in data representation, emotion labeling, and feature selection algorithm. In addition, different studies are difficult to reproduce as many of them use different public datasets or private datasets with small amounts of music clips and different levels of features.

% For computing emotional information efficiently, previous studies utilize neural networks to extract and analyze the salient semantics of acoustic features. Recent works explore neural networks given the significant improvements over hand-crafted feature-based methods \cite{piczak2015environmental,salamon2017deep,pons2019randomly,simonyan2014very}. Deep audio embedding is a type of audio features extracted by a neural network that takes audio data as an input and computes the features of the input audio. One of the main advantages of deep audio embedding representation is that spectrograms can summarize high dimensional waveforms into a compact representation. %With deep audio embedding, musical features are directly learned from the data rather than being handcrafted, i.e., human labeling. 
% Using deep audio embedding representation, 1) information can be extracted without being limited to specific kinds of data, and 2) it can save time and resources.

% \shlomo{
Previous studies have utilized neural networks to efficiently extract emotional information and analyze the salient semantics of the acoustic features. Recent works explore neural networks given the significant improvements over hand-crafted feature-based methods \cite{piczak2015environmental,salamon2017deep,pons2019randomly,simonyan2014very}. Specifically, using Convolutional Neural Networks (CNNs) and Recurrent Neural Networks (RNNs) based models, several studies attempt to extract necessary parameters for emotion prediction and reduce the dimensionality of the corresponding emotional features \cite{cheuk2020regression,thao2019multimodal,dong2019bidirectional,liu2019music}. After careful feature engineering, these methods are suitable for a target data set for emotion prediction, however, a considerable amount of training and optimization process is still required.

Deep audio embeddings are a type of audio features extracted by a neural network that take audio data as an input and compute features of the input audio. The advantages of deep audio embedding representations are that they summarize the high dimensional spectrograms into a compact representation. Using deep audio embedding representation, 1) information can be extracted without being limited to specific kinds of data, and 2) it can save time and resources.

% Several studies have conducted a music classification task using deep audio embedding methods.For example, Choi et al. implement a convet feature-based deep audio embedding and show how it can be used in six different music tagging tasks, such as dance genre classification, genre classification, speech/music classification, emotion prediction, vocal/non-vocal classification, and audio event classification~\cite{choi2017transfer}.%Especially, the emotion prediction task showed state-of-the-art performance.
% Kim et al. evaluate important factors of deep audio embeddings and propose several statistical methods to understand their usage in learning tasks. However, existing studies lack in the analysis of deep audio embeddings using multiple music dataset for music emotion recognition.

% \shlomo{
Several studies have used deep audio embedding methods in music classification tasks. For example, Choi et al. implemented a convnet feature-based deep audio embedding and showed how it can be used in six different music tagging tasks such as dance genre classification, genre classification, speech/music classification, emotion prediction, vocal/non-vocal classification, and audio event classification~\cite{choi2017transfer}. Kim et al. proposed several statistical methods to understand deep audio embeddings for usage in learning tasks \cite{kim2019nearby}. However, there are currently no studies analyzing the use of deep audio embeddings in the MER task across multiple datasets.

% Transfer learning has been shown great potential to 1) identify transferable knowledge by accommodating new knowledge, 2) retain previously learned experiences, and 3) use limited resources efficiently. Previous MIR studies demonstrate the possibility of transfer learning for deep audio embedding solutions. However, researchers have not fully investigated the efficacy of multiple deep audio embeddings for the MER task.

% In order to analyze emotional information from music, it is necessary to recognize the unique characteristics of each song. The goal of SED is to detect different sound events in audio streams. Cramer et al.~\cite{cramer2019look} propose a new audio analysis method for SED based on computer vision techniques to processing audio data. Enhancing sound features by knowledge transfer is getting attention in the Music Information Retrieval (MIR) research. The recent MIR studies report considerable performance improvements in music analysis, indexing, and classification tasks by cross-domain knowledge transfer \cite{hamel2010learning,van2013deep}. However, it further remains to be seen how deep audio embedding models can be flexible for the emotion prediction task with audio knowledge learned from different domains.

% \shlomo{

Knowledge transfer is getting increased attention in the Music Information Retrieval (MIR) research as a method to enhance sound features. Recent MIR studies report considerable performance improvements in music analysis, indexing, and classification tasks by using cross-domain knowledge transfer \cite{hamel2010learning,van2013deep}. For automatic emotion recognition in speech data, Feng and Chaspari used a Siamese neural network for optimizing pairwise differences between source and target data \cite{feng2020siamese}. In the context of SED, where the goal is to detect different sound events in audio streams, Cramer et al.~\cite{cramer2019look} propose a new audio analysis method, using deep audio embeddings, based on computer vision techniques. It remains to be seen if knowledge transfer can be successfully applied on deep audio embeddings from the SED domain to the MIR domain for the task of MER.

% \jason{why are you talking about transfer learning?}
% Transfer learning has been shown great potential to 1) identify transferable knowledge by accommodating new knowledge, 2) retain previously learned experiences, and 3) use limited resources efficiently.
% Using transfer learning, it is possible to reuse the output of a pre-trained neural network as the basis for a new learning task.
% Previous works on transfer learning in MIR mostly use intra-domain knowledge transfer~\cite{lu2018vocal,choi2017transfer}.
% For example, Lu et al. proposed the system of melody extraction based on audio-symbolic music domain transfer learning. Bittner et al. adopt a fully convolution neural network and output a salience representation at the song level~\cite{bittner2017deep}. Based on transfer learning, For cross-domain knowledge transfer in MIR, Lee et al. investigated users' music listening log data for music application tasks \cite{lee2019enhancing}.

In this study, we use deep audio embedding methods designed for the SED task and apply it over four music emotion datasets for learning emotion features in music.

\section{Methods}

\subsection{Downstream Task: Music Emotion Recognition}

% Annotating emotion annotation on music data precisely is quite challenging; it is labor-intensive and time-consuming tasks. To solve this problem, 

% We use a deep transfer learning method, which incorporates the information in audio data to assist in training the emotion prediction model.
% \jason{explain the intuition behind this proposed method.}
% \jason{formulate the problem here and then explain the method below.}

% The goal of the emotion recognition task is to predict the corresponding emotion label that indicates the annotation of emotion categories of each music clip. For our proposed approach for music emotion recognition, 
% : first, given a song as an input file, and pre-processing a deep audio embedding feature vector using a deep audio embedding method, next, selecting a classifier for classifying each deep audio embedding feature in a emotion category 

We employ a two-step experimental approach (see Figure \ref{fig:tl_workflow}). 
% In this approach, the time-series audio input (song) is represented by a deep audio embedding feature vector and the emotion category includes emotional contexts such as arousal and valence properties.

\vskip 0.05in

% Figure \ref{fig:tl_workflow} shows the two-step workflow of this paper.

Step 1. Given a song as an input, a deep audio embedding model extracts the deep audio embeddings that indicate the acoustic features of the song.
\vskip 0.03in
Step 2. After extracting deep audio embeddings, the selected classification model predicts the corresponding emotion category that indicates the emotion label of the song.

\subsection{Deep Audio Embeddings}

%For our deep transfer learning, 
% investigate deep audio embeddings and how it can be generated.
We choose two deep audio embedding methods, $L^3$-Net and VGGish, which are state-of-the-art audio representations pre-trained on 60M AudioSet \cite{gemmeke2017audio} and Youtube-8M data \cite{abu2016youtube}. AudioSet and Youtube-8M are large labeled training datasets that are widely used in audio and video learning with deep neural networks.

\begin{figure}[htbp]
  \centering
  \includegraphics[width=\columnwidth]{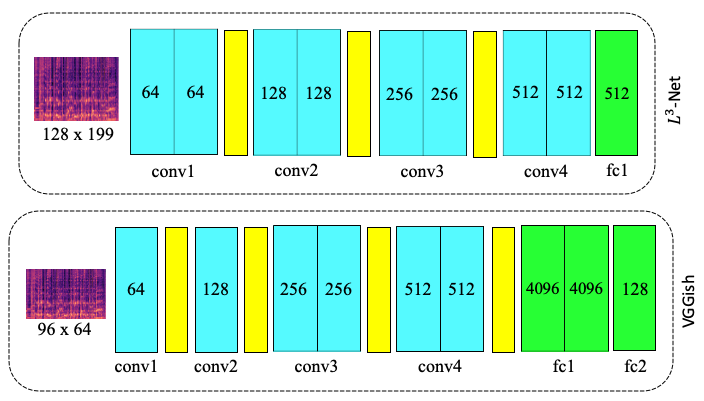}
\caption{\textbf{Network Architecture of $L^3$-Net and VGGish.} The input spectrogram representations are 128x199 for $L^3$-Net and 96x64 for VGGish. Blue boxes, yellow boxes, and green boxes denote the 2D convolutional layers, max-pooling layers, and fully-connected layers, respectively. The number inside of the blue box is the size of filters and the number inside of the green box is the number of neurons.}

\label{fig:vgg_$L^3$}
\end{figure}

\subsubsection{Look, Listen, and Learn network ($L^3$-Net)}

%Cramer et al. \cite{cramer2019look} introduce 
$L^3$-Net is an audio embedding method \cite{cramer2019look} motivated by the original work of Look, Listen, and Learn ($L^3$) \cite{arandjelovic2017look} that processes Audio-Visual Correspondence learning task in computer vision research.
The key differences between the original $L^3$ (by Arandjelović and Zisserman) and $L^3$-Net (by Cramer et al.) are (1) input data format (video vs. audio), (2) final embedding dimensionality, and (3) training sample size. 

% \vskip 0.03in

The $L^3$-Net audio embedding method consists of 2D convolutional layers and 2D max-pooling layers, and each convolution layer is followed by batch normalization and a ReLU nonlinearity (see Figure \ref{fig:vgg_$L^3$}). For the last layer, a max-pooling layer is performed to produce a single 512 dimension feature vector ($L^3$-Net serves as an option for output embedding size such as 6144 or 512, and we choose 512 as our embedding size). The $L^3$-Net method is pre-trained on Google AudioSet 60M training samples containing mostly musical performances \cite{gemmeke2017audio}.

% \vskip 0.03in
We follow the design choices of the $L^3$-Net study which result in the best performance in their SED task. We use Mel spectrograms with 256 Mel bins spanning the entire audible frequency range, resulting in a 512 dimension feature vector. We revise OpenL3 open-source implementation\footnote{\scriptsize{OpenL3 open-source library:}\scriptsize\url{https://openl3.readthedocs.io/en/latest/index.html}}%how many training data they have
for our experiments.

%In the remainder of the paper, this feature is referred to as a pre-trained $L^3$-more feature, or simply $L^3$-more feature.

% \vspace{1mm}

\subsubsection{VGGish} % VGG-like audio embedding model

% To explore the impact of using semantically informed representation, 
We also verify another deep audio embedding method, VGGish~\cite{simonyan2014very}, VGG-structure (VGGNet) based deep audio embedding model.
VGGish is a 128-dimensional audio embedding method, motivated by VGGNet \cite{simonyan2014very}, and pre-trained on a large YouTube-8M dataset~\cite{abu2016youtube}. Original VGGNet is targeting large scale image classification tasks, and VGGish is targeting extracting acoustic features from audio waveforms. The VGGish audio embedding method consists of 2D convolutional layers and 2D max-pooling layers to produce a single 128 dimension feature vector (see Figure \ref{fig:vgg_$L^3$}). We modify a VGGish open-source implementation\footnote{\scriptsize{VGGish:}\scriptsize\url{https://github.com/tensorflow/models/tree/master/research/audioset/vggish}} for our experiments.

%%% VGGish structure 설명 빠짐

% VGGish converts audio input features into a semantically meaningful, high-level 128 dimension embedding which can be fed as input to a downstream classification model. Based on the TensorFlow implementation model, we can extract input features for the audio classification model from audio waveforms and post-process the model embedding output into the same format as the released embedding features. 

%The downstream model can be shallower than usual because the VGGish embedding is more semantically compact than raw audio features. This model returns an embedding of dimensionality 128 for each embedding frame}. 

\subsection{Music Emotion Classifiers}
From the computed deep audio embeddings, we predict an emotion category corresponding to each audio vector as a multi-class classification problem.
We employ six different classification models, Support Vector Machine (SVM), Naive Bayes (NB), Random Forest (RF), Multilayer Perceptron (MLP), Convolution Neural Network (CNN), and Recurrent Neural Network (RNN). 
%\jason{You could say you share the details of those classifiers in an Appendix.}

For each classification task, we use 80\% of the data for training, 10\% for testing, and 10\% for validation.
%\jason{The standard is to split the data into training/validation/testing}
% \jason{is 15\% the standard? same as in the other paper?}
All six classification models are implemented in Scikit-learn~\cite{pedregosa2011scikit}, Keras~\cite{chollet2015keras}, and Tensorflow~\cite{abadi2016tensorflow}. In the case of MLP, CNN, and RNN classification models, we share some implementation details below.

$\bullet$ MLP: %We test on the MLP model to address the multi-class classification problem. 
We implement the MLP model with two of a single hidden layer with 512 nodes, a ReLU activation function, an output layer with a number of emotion categories, and a softmax activation function. The model is processed using the categorical cross-entropy loss function and we use Adam stochastic gradient descent \cite{kingma2014adam}. %Dropouts of 50\% are applied for the input layer and a hidden layer. 
We fit the model for 1000 training epochs with the default batch size of 32 samples and evaluate the performance at the end of each training epoch on the test dataset. %Table\ref{tab:mlp} shows the structure of the MLP model. 

% \begin{table}[t]
%     \centering
%     \caption{Table Type Styles.}
%     \begin{tabular}{ |c|c|c| } \hline
%     \multicolumn{3}{|c|}{\textbf{Table Head}} \\ \hline
%     \textbf{\textit{Table column subhead}} & \textbf{\textit{Subhead}} & \textbf{\textit{Subhead}}\\ \hline
%     {More table copy} & {} & {} \\ \hline
%     \end{tabular}
%     \label{modules}
% \end{table}

% \begin{table}[t]
% \caption{\textbf{Network Architecture of CNN Classifier.} This table contains information of $L^3$-Net audio embedding. Audio Input shape would be changed based on the type of features, such as 512 for $L^3$-Net and 128 for VGGish.}
% \vskip 0.1in
% \begin{center}
% \begin{footnotesize}
% \begin{sc}
% \setlength\tabcolsep{3pt} 
% \begin{tabular}{ |l|c|c|r| }
%  \hline
% \toprule
% Layer & Filters & Output Shape  & Activation \\ \hline
% \midrule
% Audio Input   &  &  (1, 512) & \\ \hline
% \midrule

% Conv 1D(3)& 64 & (510, 64)&ReLU\\  \hline
% Conv 1D(3)& 64 &(508, 64) &ReLU\\ \hline
% MaxPooling 1D(3) & & (169, 64)& \\ \hline
% \midrule

% Conv 1D(3) & 128 & (167, 128)  &ReLU\\  \hline
% Conv 1D(3) & 128 & (165, 128)& ReLU\\ \hline
% Global \\Average Pooling && 128& \\ \hline
% \midrule
% Dropout(0.5)& & 128&\\ \hline
% \midrule
% Dense & &\# of classes & softmax \\ \hline
% % \midrule
% % Output Layer & & \\

% % \midrule
% \bottomrule
% \end{tabular}
% \end{sc}
% \end{footnotesize}
% \end{center}
% % \vskip -0.3in

% \label{tab:cnn}
% \end{table}

% \subsubsection{Convolution Neural Networks (CNNs)}

$\bullet$ CNN: %Table \ref{tab:cnn} describes the structure of CNNs used in this paper. 
For CNN classification model, we revise the convolutional filter design proposed by Abdoli et al. \cite{abdoli2019end}, which includes four 1D convolution layers and a 1D max-pooling operation layer. Each layer processes 64 convolutional filters. The input to the network is a Mel spectrogram, size of 512 feature vector extracted from a deep audio embedding method. This input size is varied depending on the type of embedding methods. For example, in the case of $L^3$-Net, the embedding size is 512, VGGish embedding size is 128. ReLU activation functions are applied to the convolutional layers %, $f(x) = max(0,x)$, 
to reduce the backpropagation errors and accelerate the learning process \cite{goodfellow2016deep}. The softmax function is used as the output activation function with a number of emotion categories. Adam optimizer, categorical cross-entropy loss function, and the batch size of 32 samples are used. The stopping criterion is set as 1000 epochs with an early-stopping rule if there is no improvement to the score during the last 100 learning epochs.

$\bullet$ RNN: %Table \ref{tab:rnn} shows the structure of the LSTM-RNN model that is used in this paper. 
Weninger et al. \cite{weninger2014line} propose LSTM-RNN design as an automaton-like structure mapping from an observation sequence to an output feature sequence. We use LSTM networks with a pointwise softmax function based on a number of emotion categories. Adam optimizer, the categorical cross-entropy loss function, and the batch size of 32 samples are used. The same stopping criterion is set as CNNs.

% While using an RNN to explicitly model conditional dependencies between the labels, the model would be able to exploit the full audio input. 

% Based on the entries of the matrix, it is possible to compute sensitivity (recall), specificity, and precision. For a single cutoff, these quantities lead to balanced accuracy (sensitivity and specificity). 

\section{Evaluation}
\subsection{Dataset}

Four different datasets are selected for computing the emotional features in music data. In Table \ref{tab:dataset}, we show the number of music files of each dataset by emotion category.

\begin{table}[t]
\caption{\textbf{Dataset Details.} The number of emotion categories in each dataset and the number of clips in each emotion category are described. Q1, Q2, Q3, Q4 means the emotion categories of the four Arousal-Valence (A-V) quadrants based on Russell's model \cite{russell2003core}: Q1 (A+V+), Q2 (A+V-), Q3 (A-V-), Q4 (A-V+). For RAVDESS singing data, it has been classified into six emotion categories, N:Neutral, C:Calm, H:Happy, S:Sad, A:Angry, F:Fearful}

\vskip 0.1in
\begin{center}
\begin{footnotesize}
\begin{sc}
\setlength\tabcolsep{3pt} 
\begin{tabular}{ lccccr}
% \toprule
 \hline
& \multicolumn{4}{c} {Emotion Category} &                            \\
 \hline
% \midrule
Dataset & Q1 & Q2 & Q3 & Q4 & Total \\ \hline
% \midrule
4Q Audio Emotion & 225 & 225& 225 & 225 &900 \\ \hline %$\surd$ \\
Bi-modal Emotion  & 52 & 45& 31& 34 &162\\  \hline %$\times$\\
Emotion in Music & 305 & 87 & 241 &111 &744\\ \hline
% \midrule
% \bottomrule
\end{tabular}

\medskip

% \footnotesize
% \fontsize{10}
\begin{tabular}{lccccccr}
% \toprule
 \hline
& \multicolumn{6}{c} {Emotion Category} &                            \\
 \hline
% \midrule
Dataset & N & C & H & S &A &F & Total \\ \hline
% \midrule
\footnotesize{RAVDESS}& 92 & 184& 184 & 184 &184 & 20 & 848 \\  \hline
%$\surd$ \\
% \midrule
% \bottomrule

%Neutral
%Calm
%Happy
%Sad
%Angry
%Fearful
\end{tabular}
\end{sc}
\end{footnotesize}
\end{center}
% \vskip -0.3in

\label{tab:dataset}
\end{table}

$\bullet$ 4Q Audio Emotion Dataset: This dataset is introduced by Panda et al. \cite{panda2018musical}, annotated each music clip into four Arousal-Valence (A-V) quadrants based on Rusell's model \cite{russell2003core}: Q1 (A+V+), Q2 (A+V-), Q3 (A-V-), Q4 (A-V+). Each emotion category has 225 music clips, and each music clip is 30 seconds long. The total music clips for the dataset are 900 files.

$\bullet$ Bi-modal Emotion Dataset: This dataset is introduced by Malheiro et al. \cite{malheiro2016bi} in a
context of bi-modal analysis in the emotion recognition with audio and lyric information. The emotion category is also annotated into four A-V quadrants by Russell's model. In this dataset, each emotion category has a different number of music clips, Q1: 52 clips; Q2: 45 clips; Q3: 31 clips, and Q4: 34 clips, and each music clip is 30 seconds long. The total music clips for the dataset are 162 files. The size of this dataset is the smallest for our experiments. %The emotion category was annotated by 39 human subjects.

$\bullet$ Emotion in Music: Using a crowdsourcing platform, Soleymani et al. \cite{soleymani20131000} release a music emotion dataset with 20,000 arousal and valence annotations on 1,000 music clips. For our experiments, we map the arousal and valence annotation into four A-V quadrants followed by previous Russell's model settings. Each emotion category has a different number of music clips, Q1: 305 clips; Q2: 87 clips; Q3: 241 clips, and Q4: 111 clips, and each music clip is 45 seconds long. We use 744 music clips of the dataset in our experiments. This dataset is one of the most frequently used datasets for the MER task. 
% Participants listened to the sound source and judged it to create a label by classifying emotions. 
% Each emotion category was annotated by human subjects using Amazon Mechanical Turk \cite{paolacci2010running}.

$\bullet$ Ryerson Audio-Visual Database of Emotional Speech and Song (RAVDESS): This dataset is introduced by Livingstone et al. \cite{livingstone2018ryerson} for understanding the emotional context in speech and singing data. In singing data, it includes the recording clips of human singing with different emotional contexts. 24 different actors were asked to sing in six different emotional states: neutral, calm, happy, sad, angry, fearful. We choose singing data only for our experiments. Each emotion category has a different number of music clips, neutral: 92 clips; calm: 184 clips; happy: 184 clips, sad: 184 clips, angry: 184 clips, and fearful: 20 clips, and each music clip is 5 seconds long. The total music clips for the dataset are 848 files.

\subsection{Baseline Audio Features}
% \jason{Move this subsection to the Evaluation}
%MFCC embedding size
As a baseline feature, we use Mel-Frequency Cepstral Coefficients (MFCCs), which are known to be efficient low-level descriptors for timbre analysis, used as features of music tagging tasks \cite{choi2017transfer,kim2018sample}. MFCCs describe the overall shape of a spectral envelope. We first calculate the time derivatives of the given MFCCs and then take the mean and standard deviation over the time axis. Finally, we concatenate all statistics into one vector. We generate the MFCC features of each music clip into a matrix of 20 x 1500. Librosa is used for MFCCs extraction and audio processing  \cite{mcfee2015librosa}.

% MFCCs .

\subsection{Performance Measures}
% \jason{Move this to the Evaluation subsection}

For classification problems, classifier performance is typically defined according to the confusion matrix associated with the classifier. We use accuracy measure as a primary evaluation criterion. We also calculate F1-score and $r^2$ score for comparison with other baseline models.

\begin{figure}[!htb]
  \centering
  \includegraphics[width=\columnwidth]{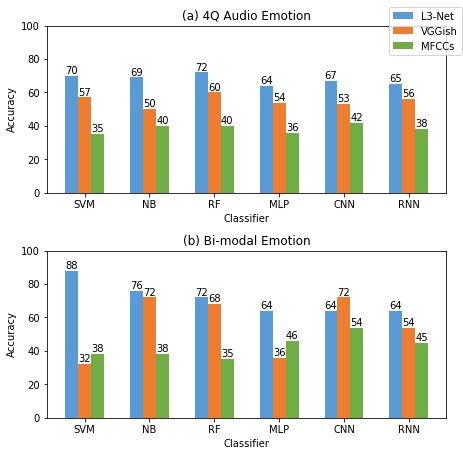}
  \includegraphics[width=\columnwidth]{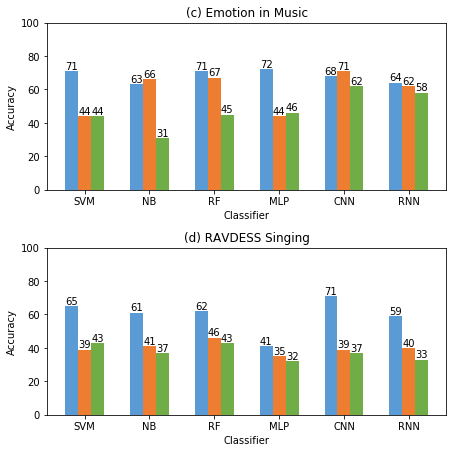}
%   \vskip 0.2in
  \caption{\textbf{Performance of Emotion Recognition on the Music Emotion Datasets.} Blue bar means the performance of $L^3$-Net, orange bar for VGGish, and green bar for MFCCs. X-axis indicates the type of classifiers we used, and Y-axis indicates the classification accuracies of the emotion category recognition.}

\label{fig:figure}
\end{figure}

\subsection{Evaluation of Music Emotion Recognition}

% The goal of this test is to assess the relevance of deep audio embedding features to MER and understand whether the embedding methods improve the current emotion recognition performance. With this in mind, 

In Figure \ref{fig:figure}, we show the performance of deep audio embeddings over four music emotion datasets. We empirically analyze deep audio embeddings in several settings against baseline MFCC features. The experiments are validated with 20 repetitions of cross-validation where we report the average results. We share key observations in the next sections.

%  (blue bar for $L^3$-Net, orange bar for VGGish, and green bar for MFCCs in the Figure \ref{fig:figure})

\subsubsection{Performance Analyzed by Features}

% MFCCs

% This performance convinces us to use pre-trained $L^3$-Net audio embedding features for the emotion recognition task. 

The $L^3$-Net embedding has the best performance in all considered cases except for two, CNN classifier accuracy both in Bi-modal Emotion and Emotion in Music dataset (see Figure \ref{fig:figure}). Even though the $L^3$-Net embedding is generally not a descriptor for any music-related tasks before, the performance convinces us to use a pre-trained $L^3$-Net audio embedding model for the MER task.

Since the direct use of the $L^3$-Net embedding shows the better performance, we also investigate more about the different embedding dimension of the $L^3$-Net and compare the performance between 512 and 6144. Interestingly, we observe decreasing results with the dimension of 6144 $L^3$ embeddings. This indicates that those extra features might not be relevant but introducing noise. While the 512 $L^3$ embeddings show consistent higher performance in many cases, based on our observations, even we increase the depth and number of parameters, 6144 $L^3$-Net embeddings perform slightly lower on this MER task. Thus, we have not included the performance in the figure. Note that reported results in Figure \ref{fig:figure} are only considered the performance of 512.
%%%%%%%%%%%%2020-11-24

% the performance of 6144 suggests that indeed 512 $L^3$-Net architecture performs slightly better on this emotion recognition task.

%The result suggests that indeed 512 $L^3$-Net architecture performs slightly better on this task. 
% except for the second figure in Figure \ref{fig:figure}. 

% The $L^3$-Net embedding method is better than the VGGish except for two cases: CNN classifier accuracy on Bi-modal Emotion data, and Emotion in Music dataset. This is a surprise, but a possible explanation is that the VGGish model with CNN

% In contrast, VGGish embeddings show higher performance in RAVDESS singing dataset (see Figure \ref{fig:figure}-(d)). Considering RAVDESS experimental results show low overall performance, it is assumed that the structure used in the experiment is not suitable for this data set. We thought of the reason as the type and the length of music clips; which is the dataset of vocal singing. It suggests that an estimated $L^3$-Net embedding contains the high-level information related to music which is not best suited for singing data yet.

Comparing between $L^3$-Net and VGGish, $L^3$-Net outperforms VGGish across the dataset. This could be because $L^3$-Net was pre-trained on both visual and audio onto the same embedded space which can include more features. The performance of VGGish is better than MFCC baseline features with the rest of the classification models, even though it has fewer parameters, 128. This justifies our use of $L^3$-Net as a main deep audio embedding choice for MER task and VGGish is for some cases.

\vskip 0.03in
% Figure \ref{fig:figure} also represents the classification accuracies predicted by different classifier choices for emotion recognition task, SVM, NB, RF, MLP, CNN, and RNN. The SVM model has a much higher performance in most of the cases since it has performed well in previous MER studies. 

It is generally known that decision trees and in our case, RF is better than other neural network based classifiers where the data comprises a large set of categories \cite{pons2019randomly}. It also deals better with dependence between variables, which might increase the error and cause some significant features to become insignificant during training. SVM uses kernel trick to solve non-linear problems whereas decision trees derive hyper-rectangles in input space to solve the problem. This is why decision trees are better for categorical data and it deals with co-linearity better than SVM. We still find that SVM outperforms RF in some cases. The reason can be that SVM deals better with margins and thus better handles outliers. Although these are tangential considerations, it seems to support the overall notion that MER is a higher level recognition problem that first needs to address the division of the data into multiple acoustic categories, also requiring the learning of a rather non-trivial partition structure within these sub-categories.

\subsubsection{Performance Analyzed by Datasets}

% The four music datasets we used have the following characteristics: the size of each dataset and the length of music clip are different. For example, in the case of 4Q Audio Emotion dataset \cite{panda2018novel}, the classification accuracies may have different result if it contains less data than Bi-modal Emotion dataset \cite{malheiro2016bi}. Also, in the case of the RAVDESS singing data, it is a 5 seconds long file recording of human singing voice without any background music, different from the three other datasets. So it may have different results if it is extracted by other deep audio embedding methods than from the existing SED task.

% In addition to the aspects of music datasets, 

% each dataset also has different classification performance in existing emotional recognition studies. 

For comparison with prior works studying emotions in audio signals, we analyze the performance of previous studies on each dataset we used. We choose four baseline MER models for our experiments: 1) Panda et al. \cite{panda2018musical} release the 4Q Music Emotion dataset and present the study of musical texture and expressivity features, 2) Malheiro et al. \cite{malheiro2016bi} present novel lyrical features for MER task and release the Bi-modal Emotion dataset, 3) Choi et al. \cite{choi2017transfer} present a pre-trained convnet feature for music classification and regression tasks and evaluate the model using Emotion in Music dataset, 4) Arora and Chaspari \cite{arora2018exploring} present the method of a siamese network for speech emotion classification and evaluate the method using RAVDESS dataset. We compare those baseline models to the performance of our proposed method (see Table \ref{tab:comparecompare}).

\begin{table}[H]
\caption{\textbf{Performance Comparison with Baseline MER models.} This table shows the data and feature information used in previous baseline models. Data column indicates each dataset for the experiment. Feature column indicates the set of feature vectors extracted by the baseline model. Metric column indicates the metric used for the performance analysis. Baseline column includes the performance of the baseline models. Proposed $L^3$-Net column includes the best performance of $L^3$-Net embeddings on each music dataset.} % (see also Table \ref{tab:f1table} and Figure \ref{fig:figure})

% Domain Knowledge means a feature set defined by domain knowledge, Convnet means features extracted by pre-trained Convnet \cite{choi2017transfer}, and Siamese means features extracted by Siamese network \cite{arora2018exploring}. F denotes the F1 score metric, $r^2$ denotes r squared score of the coefficient of determination, and ACC denotes classification accuracy. Proposed denotes the best performance of our proposed model using $L^3$-Net embeddings 

\label{tab:comparingtable}
% \vskip -0.2in
\begin{center}
\begin{footnotesize}
\begin{sc}
\setlength\tabcolsep{3pt}

\begin{tabular}{lcccr}

% \toprule
 \hline
 
%  {\scriptsize }
Data & Feature & Metric  & Baseline & \begin{tabular}[c]{@{}l@{}}Proposed \\ $L^3$-Net \end{tabular} \\ %\begin{tabular}[c]{@{}l@{}}Proposed \\Model \end{tabular} \\
 \hline
% \midrule
4Q Audio & \begin{tabular}[c]{@{}l@{}}Domain \\ Knowledge\end{tabular}  & \textit{F} & 73.5\%& 72.0\%\\ %$\surd$ 
 \hline
% \midrule
Bi-modal  & \begin{tabular}[c]{@{}l@{}}Domain \\ Knowledge\end{tabular} & \textit{F} & 72.6\% & \textbf{88.0\%}\\ %$\SI{74.0}{\percent} \pm {0.05}$
 \hline
% \midrule
EmoMusic  & Convnet & $r^2$ & \begin{tabular}[c]{@{}l@{}}A: 0.656\\V: 0.462 

\end{tabular} &
\begin{tabular}[c]{@{}l@{}}A: \textbf{0.671}\\V: \textbf{0.556}\end{tabular} \\
 \hline
% \midrule
RAVDESS  & Siamese & \textit{Acc} & 63.8\%& \textbf{71.0\%}\\
 \hline

% \toprule
% Classifier & Feature set  & Dim-of Feats& F1 score \\
% \midrule
% SVM   & Baseline& 70 & $\SI{67.5}{\percent} \pm {0.05}$\\ %$\surd$ \\ 
% %CNN=57.77
% %RNN=70.22

% SVM  & Baseline+novel& 70 & $\SI{74.0}{\percent} \pm {0.05}$\\ % $\times$\\
% SVM  & Baseline+novel & 100 & $\SI{76.0}{\percent} \pm {0.04}$\\
% \midrule

% SVM   & $L^3$ & 512 &\textbf{69.77} \\ %$\surd$ \\ 
% %CNN=57.77
% %RNN=70.22
% SVM   & VGGish & 128 &\textbf{69.77} \\

% RF   & $L^3$& 512 & 56.47\\ % $\times$\\
% RNN & $L^3$ & 512&  38.22\\
% \bottomrule
\end{tabular}
\end{sc}
\end{footnotesize}
\end{center}
% \vskip -0.3in
\label{tab:comparecompare}
\end{table}

% In Table \ref{tab:comparison}, Domain Knowledge means a feature set defined by domain knowledge in the previous studies \cite{panda2018musical,malheiro2016bi}, Convnet means features extracted by pre-trained Convnet \cite{choi2017transfer}, and Siamese means features extracted by Siamese network \cite{arora2018exploring}. F denotes the F1 score metric, $r^2$ denotes r squared score of the coefficient of determination, and ACC denotes classification accuracy.

\vskip 0.02in

In the case of the 4Q Audio Emotion dataset, the previous study by Panda et al. obtained its best result of 73.5\% F1-score with a high number of 800 features. In Table \ref{tab:comparison}, Domain Knowledge means a feature set defined by domain knowledge in the study. For achieving the performance of the previous study, the following steps are needed. First, we need to pre-process standard or baseline audio features of each audio clip. The study used Marsyas, MIR Toolbox, and PsySound3 audio frameworks to extract a total of 1702 features. Second, we need to calculate the correlation between the pair of features for normalization. After the pre-processing, the number of features can be decreased to 898 features. Third, after computing these baseline audio features, we also need to compute novel features of each audio clip proposed by the study. Those features were carefully designed based on domain expertise, such as glissando features, vibrato, and tremolo features. Finally, baseline features and extracted novel features are combined for the MER task. For the evaluation, the study conducted post-processing of the features with the ReliefF feature selection algorithm \cite{robnik2003theoretical}, ranked the features and evaluated its best-suited features. Since the performance has been evaluated by hyperparameter tuning and feature selection algorithms, these factors may influence the performance of the MER task significantly. Note that in our proposed approach, we show the performance without any post-processing.

In the case of the Bi-modal Emotion dataset, the previous study by Malheiro et al. \cite{malheiro2016bi} presented its best classification result of 72.6\% F1-score on the dataset which is lower than the performance we have, 88\% F1-score from the result of $L^3$-Net embedding with SVM classifier.

In the case of the Emotion in Music dataset, previous studies predicted the time-varying arousal and valence annotation and calculated $r^2$ score as a performance measure \cite{weninger2014line,lee2019enhancing,kim2018sample, choi2017transfer}. We previously map these time-varying annotations into four A-V quadrants based on Rusell's model and show our prediction performance with four emotion categories (see Figure \ref{fig:figure}-(c)). For a fair comparison, we also verify the original time-varying dynamic annotations from the dataset \cite{soleymani20131000} and compare the result with the baseline model. Using the Emotion in Music dataset, Choi et al. reported its $r^2$ scores of arousal annotation, 0.656 and valence annotation, 0.462 \cite{choi2017transfer}. The best performance of $L^3$-Net embeddings achieves 0.671 $r^2$ score on arousal and 0.556 $r^2$ score on valence annotation. The result shows that we have a considerable and higher performance on arousal and valence annotation. The result confirms that $L^3$-Net embedding method shows favorable performance than the previous embedding features over Emotion in Music data.

\vskip 0.01in

In the case of RAVDESS data, the study by Arora and Chaspari \cite{arora2018exploring} reported its best classification accuracy of 63.8\% over the dataset which is lower than our accuracy, 71.0\%, from the result of $L^3$-Net embedding with CNN classifier (see Figure \ref{fig:figure}-(d)).

\begin{table}[H]
\caption{\textbf{Classification Results of Each Quadrant.} The top table indicates the classification report of $L^3$-Net embedding with Random Forest classifier on 4Q Audio Emotion Dataset. The bottom table indicates the classification report of $L^3$-Net embedding with SVM classifier on Bi-modal Emotion Dataset.}
\label{tab:f1table}
% \vskip 0.2in
\begin{center}
\begin{footnotesize}

% \caption{\footnotesize (10,15)}

\begin{sc}
\setlength\tabcolsep{3pt} 
% \caption{(A) Classification Report of $L^3$-Net embedding with RF Classifier on 4Q Audio Emotion Dataset}\label{tab:1a}

\begin{tabular}{lccr}
% \toprule
 \hline
4Q Audio Emotion & Precision & Recall  & F1-score  \\ 
 \hline
% \midrule
Q1 & 0.64 & 0.85 & 0.73 \\ %$\surd$ 
 \hline
Q2 & 0.85 & 0.80 & 0.83 \\ %$\SI{74.0}{\percent} \pm {0.05}$
 \hline
Q3 & 0.73 & 0.60 & 0.66 \\ %$\surd$ 
 \hline
Q4 & 0.64 & 0.61 & 0.62 \\
 \hline
% \midrule
Accuracy  &  &  & 0.72 \\
 \hline
% \midrule
Weighted avg  & 0.73 & 0.72 & 0.72 \\
% \bottomrule
 \hline
\end{tabular}
\vskip 0.1in

\begin{tabular}{lccr}
% \toprule

 \hline
Bi-modal Emotion & Precision & Recall  & F1-score  \\ 
 \hline
% \midrule
Q1 & 0.80 & 1.00 & 0.89 \\ %$\surd$ 
 \hline
Q2 & 1.00 & 0.89 & 0.94 \\ %$\SI{74.0}{\percent} \pm {0.05}$
 \hline
Q3 & 1.00 & 0.67 & 0.80 \\ %$\surd$ 
 \hline
Q4 & 0.80 & 0.80 & 0.80 \\
 \hline
% \midrule
Accuracy  &  &  & 0.88 \\
 \hline
% \midrule
Weighted avg  & 0.90 & 0.88 & 0.88 \\
% \bottomrule
 \hline
\end{tabular}
% \vskip 0.1in
% \begin{tabular}{lccr}
% \toprule
% Emotion in Music & Precision & Recall  & F1-score  \\ 

% \midrule
% Q1 & 0.67 & 0.25 & 0.36 \\ %$\surd$ 
% Q2 & 0.75 & 0.90 & 0.82 \\ %$\SI{74.0}{\percent} \pm {0.05}$

% Q3 & 0.67 & 0.11 & 0.19 \\ %$\surd$ 
% Q4 & 0.65 & 0.82 & 0.73 \\

% \midrule
% Accuracy  &  &  & 0.71 \\
% \midrule
% Weighted avg  & 0.70 & 0.71 & 0.66 \\
% \bottomrule

% \end{tabular}

% \vskip 0.1in

% \begin{tabular}{lccr}
% \toprule
% Emotion (L3, SVM) & Precision & Recall  & F1-score  \\ 

% \midrule
% Q1 & 0.67 & 0.25 & 0.36 \\ %$\surd$ 
% Q2 & 0.75 & 0.90 & 0.82 \\ %$\SI{74.0}{\percent} \pm {0.05}$

% Q3 & 0.67 & 0.11 & 0.19 \\ %$\surd$ 
% Q4 & 0.65 & 0.82 & 0.73 \\

% \midrule
% Accuracy  &  &  & 0.71 \\
% \midrule
% Weighted avg  & 0.70 & 0.71 & 0.66 \\
% \bottomrule

% \end{tabular}
% \vskip 0.1in
% \begin{tabular}{lccr}
% \toprule
% RAVDESS (L3, SVM) & Precision & Recall  & F1-score  \\ 

% \midrule
% Neutral & 0.00 & 0.00 & 0.00 \\ %$\surd$ 
% Calm & 0.59 & 0.83 & 0.69 \\ %$\SI{74.0}{\percent} \pm {0.05}$

% Happy & 0.59 & 0.79 & 0.68 \\ %$\surd$ 
% Sad & 0.55 & 0.41 & 0.47 \\

% Angry & 0.81 & 0.83 & 0.82 \\ %$\surd$ 
% Fearful & 0.00 & 0.00 & 0.00 \\

% \midrule
% Accuracy  &  &  & 0.65 \\
% \midrule
% Weighted avg  & 0.58 & 0.65 & 0.61 \\
% \bottomrule

% \end{tabular}

\end{sc}
\end{footnotesize}
\end{center}
% \vskip -0.3in
\label{tab:comparison}
\end{table}

\subsubsection{Performance Analyzed by A-V Quadrants}

In Table \ref{tab:comparison}, we show the results analyzed by each quadrant. This classification report gives us a further understanding of the characteristic of each emotion category in music. The meaning of each quadrant (Q1, Q2, Q3, Q4) information is described in Table \ref{tab:dataset}.

In the case of the 4Q Audio Emotion dataset, Q2 and Q3 categories obtain a higher score compared to the Q1 and Q4. This indicates that emotional features in music clips with lower valence components are easier to recognize. Specifically, the Q2 category shows higher performance which is distinctive than others. Based on the dataset \cite{panda2018novel}, the study describes music clips of the Q2 category belong to specific genres, such as heavy metal, which have recognizable acoustic features than others. 
\begin{figure}[H]
  \centering
  \includegraphics[width=\columnwidth]{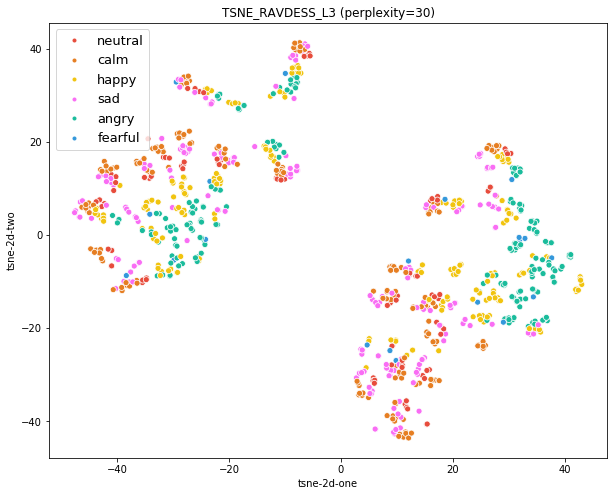}
  \includegraphics[width=\columnwidth]{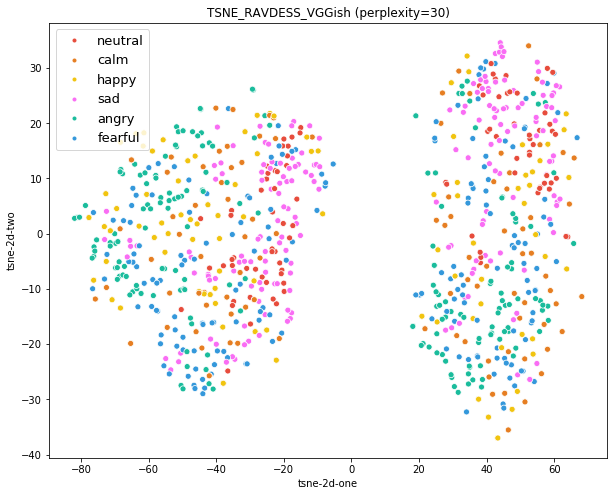}
%   \includegraphics[width=\columnwidth]{LaTeX/figs/ravdess_mfcc.png}
%   \vskip -0.2in
  \caption{\textbf{T-SNE Visualization on RAVDESS dataset.} Different colors of dots indicate the type of emotion in the dataset. Both visualizations use a perplexity value of 30. Top: T-SNE Visualization of $L^3$-Net embeddings, bottom: T-SNE Visualization of VGGish embeddings.}

\label{fig:tsne_ravdess}
\end{figure}
Lower results in Q1 and Q4 categories may also reflect the characteristics of music clips. For instance, the Q1 category indicates happy emotions, which are typically energetic based on positive arousal and positive valence components. Since Q1 and Q4 categories share the same valence axis based on Rusell's model, if the intensity of the song is not intense, the difference between the two quadrants (Q1\&Q4 or Q2\&Q3) may not be apparent. This aspect results in similar behaviors on the Q2 and Q3 categories' performances as well.

\section{Discussion and Conclusion}
\label{sec:typeset_text}

% \jason{Summarize your findings here.}
% \jason{What's the limitation of the proposed methods? How to overcome it?}

% Even though L3 not used in MER task, L3 is useful or powerful representation for MER. L3 is robust across multiple datasets outperforming other features. Overall, the result using L3 shows improvment comparing to baseline models which is statistically meaningful for bimodal, datasets.

%not just hand engineering features 
% for 4q, l3 features hand-crafted features still seems to perform better.

% In our analysis, we did not consider time evolution aspects of rhytm in 4q what they extracted. Looking into the time evolution aspects could be next step for future research.

In this paper, we evaluate $L^3$-Net and VGGish pre-trained deep audio embedding methods for MER task over 4Q Audio Emotion, Bi-modal Emotion, Emotion in Music, and RAVDESS datasets. Even though $L^3$-Net has not been intended for emotion recognition, we find that $L^3$-Net is the best representation for the MER task. Note that we achieve this performance without any additional domain knowledge feature selection method, feature training process, and fine-tuning process. Comparing to MFCC baseline features, the empirical analysis shows that $L^3$-Net is robust across multiple datasets with favorable performance. Overall, the result using $L^3$-Net shows improvement compared to baseline models for Bi-modal Emotion, Emotion in Music, and RAVDESS dataset. In the case of the 4Q Audio Emotion dataset, complex hand-crafted features (over 100 features) still seem to perform better. Specifically, our work does not consider rhythm or specific musical parameters over the time axis that 4Q Audio Emotion had, looking into time-based aspects could be the next step for future research.

%In order to gain some deeper insight 
% We use TSNE visualization. RF remove

In order to gain deeper insight into the meaning of acoustic features for emotional recognition, we use T-SNE visualization (see Figure \ref{fig:tsne_ravdess}). In both cases of $L^3$-Net and VGGish, two main clusters on the left and right side of the figure mean male/female singer groups. We can also see a relatively smooth grouping of samples by emotions with different colors. In the case of $L^3$-Net embeddings (top figure of Figure \ref{fig:tsne_ravdess}), multiple small groups in each cluster indicate individual singer which has audio recordings in different emotions. $L^3$-Net data seems to cluster into multiple smaller groups according to gender and individual categories, and this shows $L^3$-Net outperforms for detecting different timbre information than VGGish. This pattern seems to be consistent in the wild range of T-SNE perplexity parameters. This also shows that our study provides an empirical justification that $L^3$-Net outperforms VGGish, with the intuition discussed in the paper based on the clustering shown in Figure \ref{fig:tsne_ravdess}.

Accordingly, for the next step, a possible direction to validate different classifiers is to explore a combination of discrete neural learning methods, such as VQ-VAE, to first solve the categorical problem, and only later learn a more smooth decision surface. VQ-VAE has been recently explored for spectrogram-based music inpainting \cite{CSMCMuMe2020}. It would be interesting to explore similar high-level parameterization using $L^3$-Net embeddings.

{\footnotesize
\bibliography{aaai21}}

\begin{thebibliography}{48}
\providecommand{\natexlab}[1]{#1}
\providecommand{\url}[1]{\texttt{#1}}
\providecommand{\urlprefix}{URL }
\expandafter\ifx\csname urlstyle\endcsname\relax
  \providecommand{\doi}[1]{doi:\discretionary{}{}{}#1}\else
  \providecommand{\doi}{doi:\discretionary{}{}{}\begingroup
  \urlstyle{rm}\Url}\fi

\bibitem[{Abadi et~al.(2016)Abadi, Barham, Chen, Chen, Davis, Dean, Devin,
  Ghemawat, Irving, Isard et~al.}]{abadi2016tensorflow}
Abadi, M.; Barham, P.; Chen, J.; Chen, Z.; Davis, A.; Dean, J.; Devin, M.;
  Ghemawat, S.; Irving, G.; Isard, M.; et~al. 2016.
\newblock Tensorflow: A system for large-scale machine learning.
\newblock In \emph{12th $\{$USENIX$\}$ Symposium on Operating Systems Design
  and Implementation ($\{$OSDI$\}$ 16)}, 265--283.

\bibitem[{Abdoli, Cardinal, and Koerich(2019)}]{abdoli2019end}
Abdoli, S.; Cardinal, P.; and Koerich, A.~L. 2019.
\newblock End-to-end environmental sound classification using a 1d
  convolutional neural network.
\newblock \emph{Expert Systems with Applications} 136: 252--263.

\bibitem[{Abu-El-Haija et~al.(2016)Abu-El-Haija, Kothari, Lee, Natsev,
  Toderici, Varadarajan, and Vijayanarasimhan}]{abu2016youtube}
Abu-El-Haija, S.; Kothari, N.; Lee, J.; Natsev, P.; Toderici, G.; Varadarajan,
  B.; and Vijayanarasimhan, S. 2016.
\newblock Youtube-8m: A large-scale video classification benchmark.
\newblock \emph{arXiv preprint arXiv:1609.08675} .

\bibitem[{Arandjelovic and Zisserman(2017)}]{arandjelovic2017look}
Arandjelovic, R.; and Zisserman, A. 2017.
\newblock Look, listen and learn.
\newblock In \emph{Proceedings of the IEEE International Conference on Computer
  Vision}, 609--617.

\bibitem[{Arora and Chaspari(2018)}]{arora2018exploring}
Arora, P.; and Chaspari, T. 2018.
\newblock Exploring siamese neural network architectures for preserving speaker
  identity in speech emotion classification.
\newblock In \emph{Proceedings of the 4th International Workshop on Multimodal
  Analyses Enabling Artificial Agents in Human-Machine Interaction}, 15--18.

\bibitem[{Bazin et~al.(2020)Bazin, Hadjeres, Esling, and Malt}]{CSMCMuMe2020}
Bazin, T.; Hadjeres, G.; Esling, P.; and Malt, M. 2020.
\newblock Spectrogram Inpainting for Interactive Generation of Instrument
  Sounds.
\newblock
  \urlprefix\url{https://boblsturm.github.io/aimusic2020/papers/CSMC__MuMe_2020_paper_49.pdf}.

\bibitem[{Chen et~al.(2016)Chen, Zhao, Xin, Qiang, Zhang, and
  Li}]{chen2016scheme}
Chen, P.; Zhao, L.; Xin, Z.; Qiang, Y.; Zhang, M.; and Li, T. 2016.
\newblock A scheme of MIDI music emotion classification based on fuzzy theme
  extraction and neural network.
\newblock In \emph{2016 12th International Conference on Computational
  Intelligence and Security (CIS)}, 323--326. IEEE.

\bibitem[{Cheuk et~al.(2020)Cheuk, Luo, Balamurali, Roig, and
  Herremans}]{cheuk2020regression}
Cheuk, K.~W.; Luo, Y.-J.; Balamurali, B.; Roig, G.; and Herremans, D. 2020.
\newblock Regression-based music emotion prediction using triplet neural
  networks.
\newblock In \emph{2020 International Joint Conference on Neural Networks
  (IJCNN)}, 1--7. IEEE.

\bibitem[{Choi et~al.(2017)Choi, Fazekas, Sandler, and Cho}]{choi2017transfer}
Choi, K.; Fazekas, G.; Sandler, M.; and Cho, K. 2017.
\newblock Transfer learning for music classification and regression tasks.
\newblock \emph{arXiv preprint arXiv:1703.09179} .

\bibitem[{Chollet et~al.(2015)}]{chollet2015keras}
Chollet, F.; et~al. 2015.
\newblock keras.

\bibitem[{Cramer et~al.(2019)Cramer, Wu, Salamon, and Bello}]{cramer2019look}
Cramer, J.; Wu, H.-H.; Salamon, J.; and Bello, J.~P. 2019.
\newblock Look, Listen, and Learn More: Design Choices for Deep Audio
  Embeddings.
\newblock In \emph{ICASSP 2019-2019 IEEE International Conference on Acoustics,
  Speech and Signal Processing (ICASSP)}, 3852--3856. IEEE.

\bibitem[{DCASE(2019)}]{DCASE2019Task4}
DCASE. 2019.
\newblock {Detection and Classification of Acoustic Scenes and Events. Task 4:
  Sound Event Detection in Domestic Environments}.
\newblock
  \urlprefix\url{http://dcase.community/challenge2019/task-sound-event-detection-in-domestic-environments}.

\bibitem[{Dong et~al.(2019)Dong, Yang, Zhao, and Li}]{dong2019bidirectional}
Dong, Y.; Yang, X.; Zhao, X.; and Li, J. 2019.
\newblock Bidirectional Convolutional Recurrent Sparse Network (BCRSN): An
  Efficient Model for Music Emotion Recognition.
\newblock \emph{IEEE Transactions on Multimedia} 21(12): 3150--3163.

\bibitem[{Feng and Chaspari(2020)}]{feng2020siamese}
Feng, K.; and Chaspari, T. 2020.
\newblock A Siamese Neural Network with Modified Distance Loss For Transfer
  Learning in Speech Emotion Recognition.
\newblock \emph{arXiv preprint arXiv:2006.03001} .

\bibitem[{Gemmeke et~al.(2017)Gemmeke, Ellis, Freedman, Jansen, Lawrence,
  Moore, Plakal, and Ritter}]{gemmeke2017audio}
Gemmeke, J.~F.; Ellis, D.~P.; Freedman, D.; Jansen, A.; Lawrence, W.; Moore,
  R.~C.; Plakal, M.; and Ritter, M. 2017.
\newblock Audio set: An ontology and human-labeled dataset for audio events.
\newblock In \emph{2017 IEEE International Conference on Acoustics, Speech and
  Signal Processing (ICASSP)}, 776--780. IEEE.

\bibitem[{Goodfellow, Bengio, and Courville(2016)}]{goodfellow2016deep}
Goodfellow, I.; Bengio, Y.; and Courville, A. 2016.
\newblock \emph{Deep learning}.
\newblock MIT press.

\bibitem[{Hamel and Eck(2010)}]{hamel2010learning}
Hamel, P.; and Eck, D. 2010.
\newblock Learning features from music audio with deep belief networks.
\newblock In \emph{ISMIR}, volume~10, 339--344. Utrecht, The Netherlands.

\bibitem[{Hershey et~al.(2017)Hershey, Chaudhuri, Ellis, Gemmeke, Jansen,
  Moore, Plakal, Platt, Saurous, Seybold et~al.}]{hershey2017cnn}
Hershey, S.; Chaudhuri, S.; Ellis, D.~P.; Gemmeke, J.~F.; Jansen, A.; Moore,
  R.~C.; Plakal, M.; Platt, D.; Saurous, R.~A.; Seybold, B.; et~al. 2017.
\newblock CNN architectures for large-scale audio classification.
\newblock In \emph{2017 ieee international conference on acoustics, speech and
  signal processing (icassp)}, 131--135. IEEE.

\bibitem[{Jansen et~al.(2017)Jansen, Gemmeke, Ellis, Liu, Lawrence, and
  Freedman}]{jansen2017large}
Jansen, A.; Gemmeke, J.~F.; Ellis, D.~P.; Liu, X.; Lawrence, W.; and Freedman,
  D. 2017.
\newblock Large-scale audio event discovery in one million youtube videos.
\newblock In \emph{2017 IEEE International Conference on Acoustics, Speech and
  Signal Processing (ICASSP)}, 786--790. IEEE.

\bibitem[{Jansen et~al.(2018)Jansen, Plakal, Pandya, Ellis, Hershey, Liu,
  Moore, and Saurous}]{jansen2018unsupervised}
Jansen, A.; Plakal, M.; Pandya, R.; Ellis, D.~P.; Hershey, S.; Liu, J.; Moore,
  R.~C.; and Saurous, R.~A. 2018.
\newblock Unsupervised learning of semantic audio representations.
\newblock In \emph{2018 IEEE International Conference on Acoustics, Speech and
  Signal Processing (ICASSP)}, 126--130. IEEE.

\bibitem[{Kim et~al.(2019)Kim, Urbano, Liem, and Hanjalic}]{kim2019nearby}
Kim, J.; Urbano, J.; Liem, C.; and Hanjalic, A. 2019.
\newblock Are Nearby Neighbors Relatives?: Are Nearby Neighbors Relatives?:
  Testing Deep Music Embeddings.
\newblock \emph{Frontiers in Applied Mathematics and Statistics} 5: 53.

\bibitem[{Kim, Lee, and Nam(2018)}]{kim2018sample}
Kim, T.; Lee, J.; and Nam, J. 2018.
\newblock Sample-level CNN architectures for music auto-tagging using raw
  waveforms.
\newblock In \emph{2018 IEEE international conference on acoustics, speech and
  signal processing (ICASSP)}, 366--370. IEEE.

\bibitem[{Kim et~al.(2010)Kim, Schmidt, Migneco, Morton, Richardson, Scott,
  Speck, and Turnbull}]{kim2010music}
Kim, Y.~E.; Schmidt, E.~M.; Migneco, R.; Morton, B.~G.; Richardson, P.; Scott,
  J.; Speck, J.~A.; and Turnbull, D. 2010.
\newblock Music emotion recognition: A state of the art review.
\newblock In \emph{Proc. ISMIR}, volume~86, 937--952.

\bibitem[{Kingma and Ba(2014)}]{kingma2014adam}
Kingma, D.~P.; and Ba, J. 2014.
\newblock Adam: A method for stochastic optimization.
\newblock \emph{arXiv preprint arXiv:1412.6980} .

\bibitem[{Lee et~al.(2019)Lee, Lee, Park, and Lee}]{lee2019enhancing}
Lee, D.; Lee, J.; Park, J.; and Lee, K. 2019.
\newblock Enhancing music features by knowledge transfer from user-item log
  data.
\newblock In \emph{ICASSP 2019-2019 IEEE International Conference on Acoustics,
  Speech and Signal Processing (ICASSP)}, 386--390. IEEE.

\bibitem[{Lin, Chen, and Yang(2013)}]{lin2013exploration}
Lin, Y.; Chen, X.; and Yang, D. 2013.
\newblock Exploration of Music Emotion Recognition Based on MIDI.
\newblock In \emph{ISMIR}, 221--226.

\bibitem[{Liu, Fang, and Huang(2019)}]{liu2019music}
Liu, H.; Fang, Y.; and Huang, Q. 2019.
\newblock Music emotion recognition using a variant of recurrent neural
  network.
\newblock In \emph{2018 International Conference on Mathematics, Modeling,
  Simulation and Statistics Application (MMSSA 2018)}. Atlantis Press.

\bibitem[{Livingstone and Russo(2018)}]{livingstone2018ryerson}
Livingstone, S.~R.; and Russo, F.~A. 2018.
\newblock The Ryerson Audio-Visual Database of Emotional Speech and Song
  (RAVDESS): A dynamic, multimodal set of facial and vocal expressions in North
  American English.
\newblock \emph{PloS one} 13(5): e0196391.

\bibitem[{Madhok, Goel, and Garg(2018)}]{madhok2018sentimozart}
Madhok, R.; Goel, S.; and Garg, S. 2018.
\newblock SentiMozart: Music Generation based on Emotions.
\newblock In \emph{ICAART (2)}, 501--506.

\bibitem[{Malheiro et~al.(2016)Malheiro, Panda, Gomes, and
  Paiva}]{malheiro2016bi}
Malheiro, R.; Panda, R.; Gomes, P.; and Paiva, R. 2016.
\newblock Bi-modal music emotion recognition: Novel lyrical features and
  dataset.
\newblock 9th International Workshop on Music and Machine
  Learning--MML’2016--in~….

\bibitem[{McFee et~al.(2015)McFee, Raffel, Liang, Ellis, McVicar, Battenberg,
  and Nieto}]{mcfee2015librosa}
McFee, B.; Raffel, C.; Liang, D.; Ellis, D.~P.; McVicar, M.; Battenberg, E.;
  and Nieto, O. 2015.
\newblock librosa: Audio and music signal analysis in python.
\newblock In \emph{Proceedings of the 14th python in science conference},
  volume~8.

\bibitem[{Panda, Malheiro, and Paiva(2018{\natexlab{a}})}]{panda2018musical}
Panda, R.; Malheiro, R.; and Paiva, R.~P. 2018{\natexlab{a}}.
\newblock Musical Texture and Expressivity Features for Music Emotion
  Recognition.
\newblock In \emph{ISMIR}, 383--391.

\bibitem[{Panda, Malheiro, and Paiva(2018{\natexlab{b}})}]{panda2018novel}
Panda, R.; Malheiro, R.~M.; and Paiva, R.~P. 2018{\natexlab{b}}.
\newblock Novel audio features for music emotion recognition.
\newblock \emph{IEEE Transactions on Affective Computing} .

\bibitem[{Pedregosa et~al.(2011)Pedregosa, Varoquaux, Gramfort, Michel,
  Thirion, Grisel, Blondel, Prettenhofer, Weiss, Dubourg
  et~al.}]{pedregosa2011scikit}
Pedregosa, F.; Varoquaux, G.; Gramfort, A.; Michel, V.; Thirion, B.; Grisel,
  O.; Blondel, M.; Prettenhofer, P.; Weiss, R.; Dubourg, V.; et~al. 2011.
\newblock Scikit-learn: Machine learning in Python.
\newblock \emph{the Journal of machine Learning research} 12: 2825--2830.

\bibitem[{Piczak(2015)}]{piczak2015environmental}
Piczak, K.~J. 2015.
\newblock Environmental sound classification with convolutional neural
  networks.
\newblock In \emph{2015 IEEE 25th International Workshop on Machine Learning
  for Signal Processing (MLSP)}, 1--6. IEEE.

\bibitem[{Pons and Serra(2019)}]{pons2019randomly}
Pons, J.; and Serra, X. 2019.
\newblock Randomly weighted CNNs for (music) audio classification.
\newblock In \emph{ICASSP 2019-2019 IEEE International Conference on Acoustics,
  Speech and Signal Processing (ICASSP)}, 336--340. IEEE.

\bibitem[{Robnik-{\v{S}}ikonja and Kononenko(2003)}]{robnik2003theoretical}
Robnik-{\v{S}}ikonja, M.; and Kononenko, I. 2003.
\newblock Theoretical and empirical analysis of ReliefF and RReliefF.
\newblock \emph{Machine learning} 53(1-2): 23--69.

\bibitem[{Russell(2003)}]{russell2003core}
Russell, J.~A. 2003.
\newblock Core affect and the psychological construction of emotion.
\newblock \emph{Psychological review} 110(1): 145.

\bibitem[{Salamon and Bello(2017)}]{salamon2017deep}
Salamon, J.; and Bello, J.~P. 2017.
\newblock Deep convolutional neural networks and data augmentation for
  environmental sound classification.
\newblock \emph{IEEE Signal Processing Letters} 24(3): 279--283.

\bibitem[{Simonyan and Zisserman(2014)}]{simonyan2014very}
Simonyan, K.; and Zisserman, A. 2014.
\newblock Very deep convolutional networks for large-scale image recognition.
\newblock \emph{arXiv preprint arXiv:1409.1556} .

\bibitem[{Soleymani et~al.(2013)Soleymani, Caro, Schmidt, Sha, and
  Yang}]{soleymani20131000}
Soleymani, M.; Caro, M.~N.; Schmidt, E.~M.; Sha, C.-Y.; and Yang, Y.-H. 2013.
\newblock 1000 songs for emotional analysis of music.
\newblock In \emph{Proceedings of the 2nd ACM international workshop on
  Crowdsourcing for multimedia}, 1--6. ACM.

\bibitem[{Thao, Herremans, and Roig(2019)}]{thao2019multimodal}
Thao, H. T.~P.; Herremans, D.; and Roig, G. 2019.
\newblock Multimodal Deep Models for Predicting Affective Responses Evoked by
  Movies.
\newblock In \emph{2019 IEEE/CVF International Conference on Computer Vision
  Workshop (ICCVW)}, 1618--1627. IEEE.

\bibitem[{Van~den Oord, Dieleman, and Schrauwen(2013)}]{van2013deep}
Van~den Oord, A.; Dieleman, S.; and Schrauwen, B. 2013.
\newblock Deep content-based music recommendation.
\newblock In \emph{Advances in neural information processing systems},
  2643--2651.

\bibitem[{Wang, Wang, and Lanckriet(2015)}]{wang2015histogram}
Wang, J.-C.; Wang, H.-M.; and Lanckriet, G. 2015.
\newblock A histogram density modeling approach to music emotion recognition.
\newblock In \emph{2015 IEEE International Conference on Acoustics, Speech and
  Signal Processing (ICASSP)}, 698--702. IEEE.

\bibitem[{Weninger, Eyben, and Schuller(2014)}]{weninger2014line}
Weninger, F.; Eyben, F.; and Schuller, B. 2014.
\newblock On-line continuous-time music mood regression with deep recurrent
  neural networks.
\newblock In \emph{2014 IEEE International Conference on Acoustics, Speech and
  Signal Processing (ICASSP)}, 5412--5416. IEEE.

\bibitem[{Wilkinghoff(2020)}]{wilkinghoff2020open}
Wilkinghoff, K. 2020.
\newblock On open-set classification with L3-Net embeddings for machine
  listening applications.
\newblock In \emph{28th European Signal Processing Conference (EUSIPCO)}.

\bibitem[{Yang, Dong, and Li(2018)}]{yang2018review}
Yang, X.; Dong, Y.; and Li, J. 2018.
\newblock Review of data features-based music emotion recognition methods.
\newblock \emph{Multimedia Systems} 24(4): 365--389.

\bibitem[{Yang and Chen(2012)}]{yang2012machine}
Yang, Y.-H.; and Chen, H.~H. 2012.
\newblock Machine recognition of music emotion: A review.
\newblock \emph{ACM Transactions on Intelligent Systems and Technology (TIST)}
  3(3): 1--30.

\end{thebibliography}

\end{document}